\documentclass[12pt]{iopart}
\usepackage{iopams}  

\usepackage{braket} 
\usepackage{graphicx}
\usepackage{bm}

\begin{document}

\title{Asymptotic Limits of the Wigner $15J$-Symbol with Small Quantum Numbers}

\author{Liang Yu}

\address{Department of Physics, University of California, Berkeley, California 94720 USA}
\ead{liangyu@wigner.berkeley.edu}

\begin{abstract}
We present new asymptotic formulas for the Wigner $15j$-symbol with two, three, or four small quantum numbers, and provide numerical evidence of their validity. These formulas are of the WKB form and are of a similar nature as the Ponzano-Regge formula for the Wigner $6j$-symbol. They are expressed in terms of edge lengths and angles of geometrical figures associated with angular momentum vectors. In particular, the formulas for the $15j$-symbol with two, three, and four small quantum numbers are based on the geometric figures of the $9j$-, $6j$-, and $3j$-symbols, respectively, The geometric nature of these new asymptotic formulas pave the way for further analysis of the semiclassical limits of vertex amplitudes in loop quantum gravity models.
\end{abstract}


\section{Introduction}

The asymptotic limits of the Wigner $3nj$-symbols have played an important role in finding the semiclassical limits of the simplicial models of quantum gravity. The earliest example is the observation by Ponzano and Regge \cite{ponzano-regge-1968} that the semiclassical limit of the Wigner $6j$-symbol reproduces the Einstein-Hilbert action in Regge's discrete formulation of three-dimensional gravity. In more recent models of four-dimensional quantum gravity, the Wigner $15j$-symbol forms the building blocks for the vertex amplitudes of a $4$-simplex in the Barrett-Crane model \cite{barrett1998, barrett2000}, the Freidel-Krasnov model \cite{conrady2008, freidel2003}, and the EPRL model \cite{eprl2008}. See the more recent papers \cite{barrett2009, barrett2010} for discussions on the asymptotic limits of the $15j$-symbol and the classical limits of quantum gravity models. 

In this paper, we present new asymptotic formulas for the Wigner $15j$-symbol with two, three, or four small quantum numbers, when the remaining quantum numbers are taken large. The derivations of these formulas are generalizations of the derivations presented in Refs.\ \cite{yu2011a, yu2011b}. The generalizations involve the addition of extra spinors and some minor relabeling of the angular momentum vectors. We omit the details of the derivations, and only present the asymptotic formulas, and provide numerical plots of our formulas to support their validity. 

An interesting aspect of these formulas is that they present a sort of reduction of the $15j$-symbol into lower $3nj$-symbols. Although the formulas do not decompose the $15j$-symbol directly into a product of two lower $3nj$-symbols, the asymptotic limits are in terms of vector diagrams associated with lower $3nj$-symbols. For example, with four small quantum numbers, the $15j$-symbol has an asymptotic limit expressed in terms of the geometry of a triangle, which is the vector diagram for a $3j$-symbol. With three small quantum numbers, the asymptotic limit of the $15j$-symbol is expressed in terms of the geometry of a tetrahedron, which is the vector diagram for a $6j$-symbol.

Some of the basic references of the $15j$-symbol include the texts on angular momentum theory \cite{biedenharn1981, edmonds1960, varshalovich1981}. The semiclassical analysis of the $3j$-, $6j$-, and $9j$-symbols when all quantum numbers are large are given in Refs.\  \cite{littlejohn2010b, littlejohn2007, littlejohn2010a}, which form the basis for the semiclassical analysis of the $9j$-, $12j$-symbols with small and large quantum numbers \cite{yu2011a, yu2011b}. The last two papers develop and illustrate the method used to derive the formulas in this paper, so we assume familiarity with them.

The outline of this paper is as follows: In section \ref{ch8: sec_spin_network}, we present the definition of the $15j$-symbol in terms of its spin-network, and relate it to the pentagon diagram used to describe the $10j$-symbol in loop quantum gravity.  In section \ref{ch8: sec_derivation}, we briefly summarize the method of derivation for the formulas presented in this paper. We present the asymptotic formulas for the $15j$-symbol with two, three, or four small quantum numbers in sections \ref{ch8: sec_15j_2_13_case}, \ref{ch8: sec_15j_3_12_case}, and \ref{ch8: sec_15j_4_11_case}, respectively. The last section contains comments and discussions.

\section{\label{ch8: sec_spin_network}The Spin-Network for the Wigner $15J$-Symbol}

\begin{figure}[tbhp]
\begin{center}
\includegraphics[width=0.50 \textwidth]{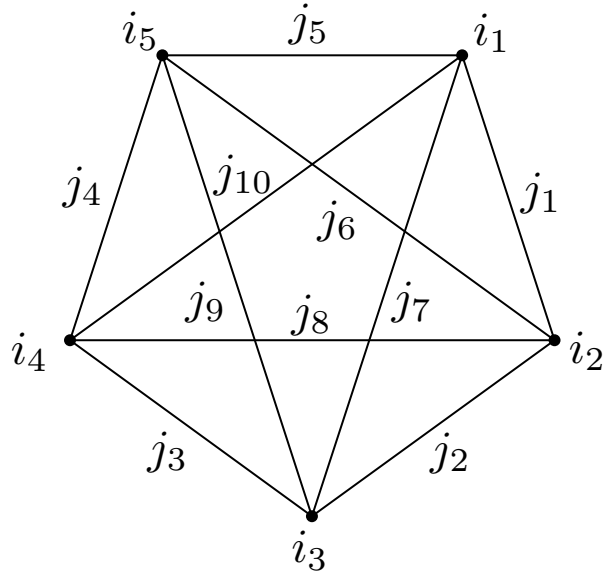}
\caption{The pentagon diagram representing the $4$-simplex amplitude in the spin foam models in four dimensions.}
\label{ch8: fig_pentagon_diagram1}
\end{center}
\end{figure}

We briefly discuss where the Wigner $15j$-symbol appear in the models of loop quantum gravity. The graphical representation of a vertex amplitude in the quantum gravity literature has the form of a pentagon diagram, which is the $SO(4)$ spin network dual to a $4$-simplex. This diagram is displayed in Fig.\ \ref{ch8: fig_pentagon_diagram1}, where the vertices $i_r$, $r = 1, \dots, 5$, represent $SO(4)$ intertwiners, and the $j_i$, $i = 1, \dots, 10$ label the balanced representations of $SO(4)$. 

\begin{figure}[tbhp]
\begin{center}
\includegraphics[width=0.80 \textwidth]{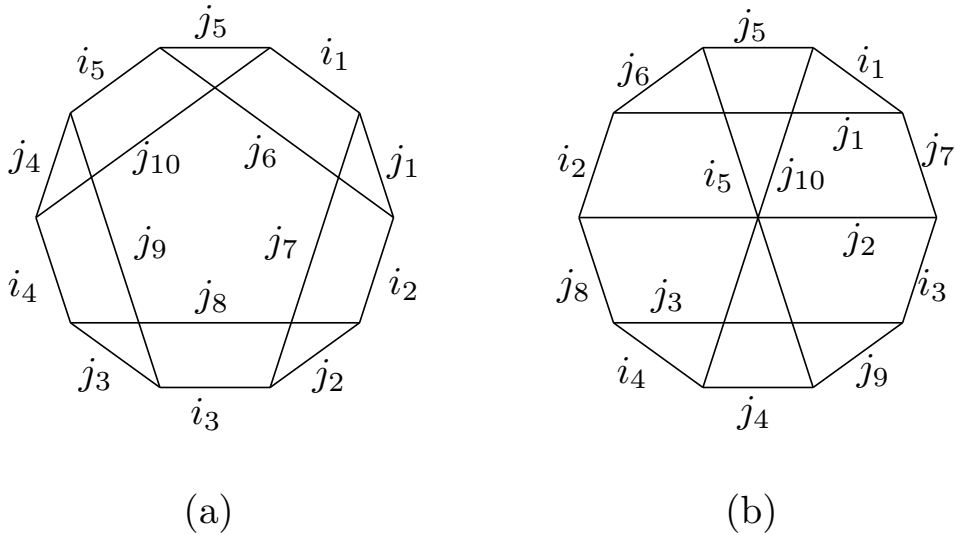}
\caption{Expanding the four-valent intertwiners in the pentagon diagram to get (a) an $SU(2)$ spin network, which is equivalent to (b) a spin network in the shape of a M\"{o}bius strip.}
\label{ch8: fig_pentagon_diagram2}
\end{center}
\end{figure}

By picking a suitable basis for the $SO(4)$ intertwiners $i_r$, $r = 1, \dots, 5$, in Fig.\ \ref{ch8: fig_pentagon_diagram1}, we can expand the pentagon diagram into a $SU(2)$ spin network with $15$ edges, which is illustrated in part (a) of Fig.\ \ref{ch8: fig_pentagon_diagram2}. This spin network is rearranged into the form of a M\"{o}bius strip in part (b) of that figure. This M\"{o}bius strip is the $SU(2)$ spin network for the Wigner $15j$-symbol of the first kind. See Figure 20.2b and Figure 20.2c in Yutsis \cite{yutsis1962}. In the models of four-dimensional quantum gravity, the vertex amplitude is denoted by a $10j$-symbol, where a product of two $SU(2)$ $15j$-symbol is summed over the intertwiners $i_r$, $r = 1, \dots, 5$. Thus, the Wigner $15j$-symbol forms the building block of the vertex amplitude of quantum gravity.

\newpage

\section{\label{ch8: sec_derivation}Methods of Derivation}

The method of derivation for the formulas in the following sections is based on multicomponent WKB theory, which is developed for the semiclassical analysis of the Wigner $3nj$-symbols with small and large quantum numbers in Refs.\  \cite{yu2011a, yu2011b}. In this section, we briefly summarize the method in those papers.

First, a $3nj$-symbol is expressed as an inner product of two wave-functions that are eigenfunctions of two sets of operators. The operators associated with each large quantum number is modeled by the Schwinger representation of angular momentum \cite{schwinger1952}. The operators associated with each small quantum number is modeled by the usual matrix representation of $SU(2)$. As a result, the eigenfunctions are multicomponent wave-functions. Following the techniques developed in Ref.\  \cite{yu2011a}, we then calculate the asymptotic form of these multicomponent wave-functions in terms of the Lagrangian manifolds associated with the large quantum numbers, and in terms of spinor fields on these Lagrangian manifolds. Finally, taking the inner product of these asymptotic wave-functions, we perform semiclassical analysis on the Lagrangian manifolds and calculate the spinor products to derive asymptotic formulas for the $3nj$-symbol.  

The application of this method is illustrated in Ref.\  \cite{yu2011a}, where it is used to derive asymptotic formulas for the $9j$-symbol with one small quantum number. The analysis contained in Ref.\  \cite{yu2011a} can be generalized to apply to the $15j$-symbol with three small quantum numbers, which will be further explained in section \ref{ch8: sec_15j_3_12_case} below. The only generalizations required include two additional spinor factors for each wave-function, and relabeling of the quantum numbers. Similarly, the analysis contained in Ref.\  \cite{yu2011b} can be generalized to apply to the $15j$-symbol with two small quantum numbers, which will be further explained in section \ref{ch8: sec_15j_2_13_case} below. Readers interested in the details of the derivations are advised to consult Refs.\  \cite{yu2011a, yu2011b}. We now present the main results of this paper, namely, the asymptotic formulas for the Wigner $15j$-symbol with small and large quantum numbers.

\section{\label{ch8: sec_15j_2_13_case}Two Small Quantum Numbers}

We will use the definition on page 66 in Ref.\ \cite{yutsis1962} for the Wigner $15j$-symbol. We will not try to cover all possible cases here, so we will make an arbitrary choice of the placement of the two small quantum numbers, and take $j_5 = s_5$ and $j_6=s_6$ to be small. The $15j$-symbol is given by an inner product of two multicomponent wave-functions, as follows:

\begin{eqnarray}
&& \left\{
   \begin{array}{ccccc}
    j_1 & j_2 & j_{12} & j_{125} & j_{1256} \\ 
    j_3 & j_4 & j_{34} &  j_{135}  & j_{1356} \\ 
    j_{13} & j_{24} & s_5 & s_6  & j_7  \\
  \end{array} 
  \right \}     \\  \nonumber
&=& \frac{\braket{ b  |  a } }{ \{ [j_{12}][j_{34}][j_{13}][j_{24}] [j_{125}] [j_{135}] [j_{1256}] [j_{1356}] \}^{\frac{1}{2}}}  \, , 
\end{eqnarray}
where the notation $[\cdot]$ denotes $[k] = 2k+1$, and

\begin{equation}
\ket{a} =  \left| 
\begin{array} { @{\,}c@{\,}c@{\,}c@{\,}c@{\;}c@{\;}c@{\,}c@{\;}c@{\;}c@{\;}c@{\;}c@{\;}c@{}}
	\hat{I}_1 & \hat{I}_2 & I_3 & \hat{I}_4 &  {\bf S}_5^2 & {\bf S}_6^2 &  \hat{I}_7 & \hat{\bf J}_{12}^2 & \hat{\bf J}_{34}^2 & \hat{\bf J}_{125}^2 &  \hat{\bf J}_{1256}^2 &  \hat{\bf J}_{{\rm tot}}  \\
	j_1 & j_2 & j_3 & j_4 & s_5 & s_6 & j_7 & j_{12} & j_{34} & j_{125} & j_{1256} & {\bf 0} 
\end{array}  \right>  \, , 
\label{ch8: eq_15j_2_13_a_state}
\end{equation}

\begin{equation}
\ket{b} =  \left| 
\begin{array} { @{\,}c@{\,}c@{\,}c@{\,}c@{\;}c@{\;}c@{\,}c@{\;}c@{\;}c@{\;}c@{\;}c@{\;}c@{}}
	\hat{I}_1 & \hat{I}_2 & I_3 & \hat{I}_4 &  {\bf S}_5^2 & {\bf S}_6^2 &  \hat{I}_7 & \hat{\bf J}_{13}^2 & \hat{\bf J}_{24}^2 & \hat{\bf J}_{135}^2 &  \hat{\bf J}_{1356}^2 &  \hat{\bf J}_{{\rm tot}}  \\
	j_1 & j_2 & j_3 & j_4 & s_5 & s_6 & j_7 & j_{13} & j_{24} & j_{135} & j_{1356} & {\bf 0} 
\end{array}  \right>  \, . 
\label{ch8: eq_15j_2_13_b_state}
\end{equation}

In the above notation, the large ket lists the operators on the top row, and the corresponding quantum numbers are listed on the bottom row. See Eq.\ (5) - (11) in \cite{yu2011b} for the definitions of some of the operators. The only ones that require new definitions are

\begin{eqnarray}
({\hat{J}}_{1256}^2)_{\alpha \beta} &=& [  J_{12}^2+ \hbar^2 s_5(s_5+1) + \hbar^2 s_6(s_6+1) ] \delta_{\alpha \beta} 
 + 2 {\bf \hat{J}}_{12} \cdot [ ({\bf S}_5)_{\alpha \beta} + ({\bf S}_6)_{\alpha \beta} ] ,  \nonumber  \\  
 && \label{ch8: eq_J1256_square} \\
({\hat{J}}_{1356}^2)_{\alpha \beta}  &=& [  J_{13}^2 + \hbar^2 s_5(s_5+1) + \hbar^2 s_6(s_6+1) ] \delta_{\alpha \beta} 
+ 2 {\bf \hat{J}}_{13} \cdot [ ({\bf S}_5)_{\alpha \beta} +  ({\bf S}_6)_{\alpha \beta} ] ,  \nonumber  \\  
 && \label{ch8: eq_J1356_square} \\
\label{ch8: eq_Jtot_vector}
({\bf \hat{J}}_{{\rm tot}})_{\alpha \beta}  &=& ( {\bf \hat{J}}_1 + {\bf \hat{J}}_2 + {\bf \hat{J}}_3 + {\bf \hat{J}}_4 + {\bf \hat{J}}_7 ) \delta_{\alpha \beta} + \hbar \, ({\bf S}_5)_{\alpha \beta} + \hbar \, ({\bf S}_6)_{\alpha \beta}.  
\end{eqnarray}

By comparing Eq.\ (11) in \cite{yu2011b} with Eq.\ (\ref{ch8: eq_Jtot_vector}) above, we find that the role of ${\bf \hat{J}}_6$ in Ref.\ \cite{yu2011b} is played by ${\bf \hat{J}}_7$ here. Otherwise, the operators associated with the large quantum numbers are exactly the same. The addition of the spinor ${\bf S}_6$ will introduce an extra spinor product in the form of a $d$-matrix, as well as additional correction terms to the phase in the asymptotic formula. 

Following the derivations for the $12j$-symbol with one small quantum number in \cite{yu2011b}, we have derived  an asymptotic formula for the $15j$-symbol with two small quantum numbers. The result is

\begin{eqnarray}
\label{ch8: eq_main_formula_15j_2_13}
&& \left\{
   \begin{array}{ccccc}
    j_1 & j_2 & j_{12} & j_{125} & j_{1256} \\ 
    j_3 & j_4 & j_{34} &  j_{135}  & j_{1356} \\ 
    j_{13} & j_{24} & s_5 & s_6  & j_7  \\
  \end{array} 
  \right \}     \\  \nonumber
&=&  \frac{(-1)^{\mu_5+\mu_6}}{  4 \pi \,  \sqrt{ (2j_{125}+1)(2j_{135}+1) (2j_{1256}+1)(2j_{1356}+1) }} \,  \\  \nonumber
&& \quad \left[  \frac{d^{s_5}_{\nu_5 \, \mu_5} (\theta^{(1)})  \, d^{s_6}_{\nu_6 \, \mu_6} (\theta^{(1)}) }{ \sqrt{| V_{123}^{(1)} V_{432}^{(1)} - V_{214}^{(1)} V_{341}^{(1)} | } } \cos \left(S^{(1)}  + (\mu_5+\mu_6) \phi_{12}^{(1)} + ( \nu_5 + \nu_6) \phi_{13}^{(1)} \right)  \right.  \\  \nonumber
&&  \left.  \quad \quad + \frac{d^{s_5}_{\nu_5 \, \mu_5} (\theta^{(2)}) \, d^{s_6}_{\nu_6 \, \mu_6} (\theta^{(2)}) }{ \sqrt{| V_{123}^{(2)} V_{432}^{(2)} - V_{214}^{(2)} V_{341}^{(2)} |}}  \sin \left(S^{(2)}  + ( \mu_5 + \mu_6) \phi_{12}^{(2)} + (\nu_5 + \nu_6) \phi_{13}^{(2)} \right)  \right]  \, ,
\end{eqnarray}
This formula is very similar to Eq.\ (80) in \cite{yu2011b}, except for an additional $d$-matrix associated with ${\bf S}_6$, and some extra phase corrections proportional to $\mu_6$ and $\nu_6$ in the arguments of the cosine and sine.

\begin{figure}[tbhp]
\begin{center}
\includegraphics[width=0.65\textwidth]{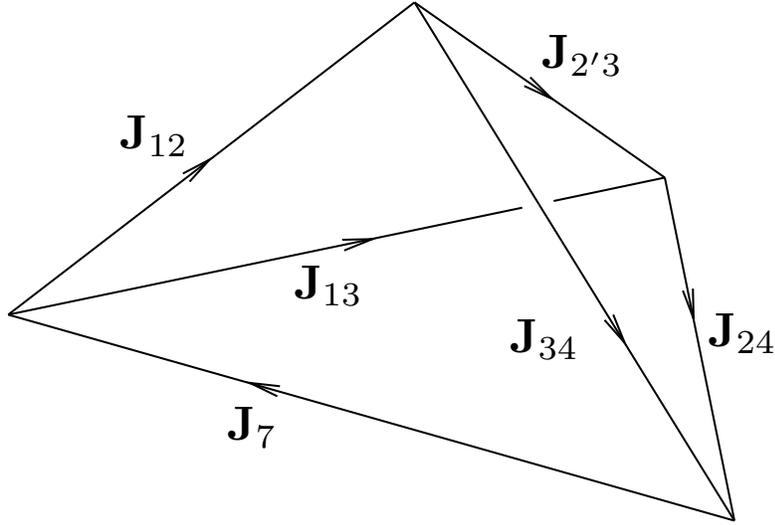}
\caption{The angles $\phi_{12}$ and $\phi_{13}$ are internal dihedral angles in the tetrahedron with the six edges ${\bf J}_7, {\bf J}_{12}, {\bf J}_{34}, {\bf J}_{13}, {\bf J}_{24}$, and ${\bf J}_{2'3}$,  where ${\bf J}_{2'3} = {\bf J}_3 - {\bf J}_2$. The angle $\theta$ is the angle between ${\bf J}_{12}$ and ${\bf J}_{13}$.}
\label{ch8: fig_tetrahedra_15j_2_13}
\end{center}
\end{figure}

In Eq.\ (\ref{ch8: eq_main_formula_15j_2_13}), the indices on the $d$-matrices are $\mu_5 = j_{125}-j_{12}$, $\nu_5 = j_{135}-j_{13}$, $\mu_6 = j_{1256} - j_{125}$, and $\nu_6 = j_{1356} - j_{135}$. The phases $S^{(1)}$ and $S^{(2)}$ are the actions associated with the $9j$-symbol, defined in Eq.\ (65), (66) in Ref.\ \cite{yu2011b}, except that ${\bf J}_6$ in \cite{yu2011b} should be replaced by ${\bf J}_7$ here. The $V$'s are given by

\begin{equation}
V_{ijk} = {\bf J}_i \cdot ({\bf J}_j \times {\bf J}_k) \,  .
\end{equation}
The angles $\phi_{12}$ and $\phi_{13}$ are internal dihedral angles at the edges $J_{12}$ and $J_{13}$, respectively, of a tetrahedron formed by the six vectors ${\bf J}_{12}, {\bf J}_{13}, {\bf J}_{24}, {\bf J}_{34}, {\bf J}_6$, and ${\bf J}_{2'3}$, where ${\bf J}_{2'3} = {\bf J}_3 - {\bf J}_2$, and $J_i = j_i + 1/2$, $i = 1,2,3,4, 7,12,34,13,24$. This tetrahedron is illustrated in Fig.\ \ref{ch8: fig_tetrahedra_15j_2_13}. See Ref.\ \cite{yu2011b} for more details on the construction of this tetrahedron. The angle $\theta$ is the angle between the vectors ${\bf J}_{12}$ and ${\bf J}_{13}$. Explicitly, the angles $\phi_{12}$, $\phi_{13}$, and $\theta$ are given by 

\begin{eqnarray}
\cos \phi_{12} &=&  \pi -  \frac{ ({\bf J}_{12} \times {\bf J}_{13} ) \cdot ({\bf J}_{12} \times {\bf J}_{7} ) }{ | {\bf J}_{12} \times {\bf J}_{13} | \,  | {\bf J}_{12} \times {\bf J}_{7}  |} \, ,    \\
\cos \phi_{13} &=&  \pi -  \frac{ ({\bf J}_{13} \times {\bf J}_{12} ) \cdot ({\bf J}_{13} \times {\bf J}_{7} ) }{ | {\bf J}_{13} \times {\bf J}_{12} | \,  | {\bf J}_{13} \times {\bf J}_{7}  |}   \, ,   \\
\cos \theta &=& \frac{ {\bf J}_{12} \cdot {\bf J}_{13} }{J_{12} J_{13} }    \, . 
\end{eqnarray}

\begin{figure}[tbhp]
\begin{center}
\includegraphics[width=0.85 \textwidth]{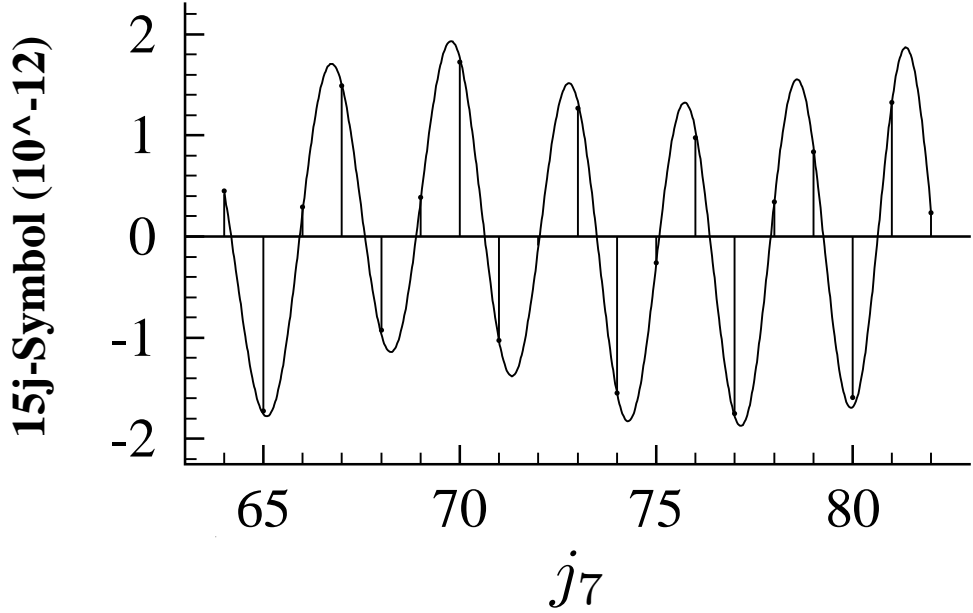}
\caption{Comparison of the exact $15j$-symbol (vertical sticks and dots) and the asymptotic formula (\ref{ch8: eq_main_formula_15j_2_13}), for the values of $j$'s shown in (\ref{ch8: eq_15j_values_2_13}). }
\label{ch8: fig_plot_15j_2_13}
\end{center}
\end{figure}

We plot the exact values of the $15j$-symbol against our approximation (\ref{ch8: eq_main_formula_15j_2_13}) in Fig.\ \ref{ch8: fig_plot_15j_2_13} for the following values of the $j$'s: 

\begin{equation}
\label{ch8: eq_15j_values_2_13}
 \left\{
  \begin{array}{ccccc}
    j_1 & j_2 & j_{12} & j_{125} & j_{1256} \\ 
    j_3 & j_4 & j_{34} &  j_{135}  & j_{1356} \\ 
    j_{13} & j_{24} & s_5 & s_6  & j_7  \\
  \end{array} 
  \right\}   
= 
	\left\{
  \begin{array}{rrrrr}
    197/2 & 187/2 & 74 & 75 & 74 \\ 
    173/2 & 205/2 & 88 & 96 & 97  \\ 
    95 & 90 & 1 & 1 & j_7 \\
  \end{array} 
  \right\}  \, . 
\end{equation}
The agreement is excellent. 

\newpage

\section{\label{ch8: sec_15j_3_12_case}Three Small Quantum Numbers}

We now treat the case where three quantum numbers are small. We take $j_3 = s_3$, $j_5=s_5$, and $j_6=s_6$ to be small. As before, we write the $15j$-symbol as a scalar product of two multicomponent wave-functions,

\begin{eqnarray}
&& \left\{
   \begin{array}{ccccc}
    j_1 & j_2 & j_{12} & j_{125} & j_{1256} \\ 
    s_3 & j_4 & j_{34} &  j_{135}  & j_{1356} \\ 
    j_{13} & j_{24} & s_5 & s_6  & j_7  \\
  \end{array} 
  \right \}     \\  \nonumber
&=& \frac{\braket{ b  |  a } }{ \{ [j_{12}][j_{34}][j_{13}][j_{24}] [j_{125}] [j_{135}] [j_{1256}] [j_{1356}] \}^{\frac{1}{2}}}  \, , 
\end{eqnarray}
where

\begin{equation}
\ket{a} =  \left| 
\begin{array} { @{\,}c@{\;}c@{\;}c@{\,}c@{\;}c@{\;}c@{\,}c@{\;}c@{\;}c@{\;}c@{\;}c@{\;}c@{}}
	\hat{I}_1 & \hat{I}_2 & {\bf S}_3^2 & \hat{I}_4 &  {\bf S}_5^2 & {\bf S}_6^2 &  \hat{I}_7 & \hat{\bf J}_{13}^2 & \hat{\bf J}_{24}^2 & \hat{\bf J}_{135}^2 &  \hat{\bf J}_{1356}^2 &  \hat{\bf J}_{{\rm tot}}  \\
	j_1 & j_2 & s_3 & j_4 & s_5 & s_6 & j_7 & j_{13} & j_{24} & j_{135} & j_{1356} & {\bf 0} 
\end{array}  \right>  \, ,
\label{ch8: eq_15j_3_12_a_state}
\end{equation}

\begin{equation}
\ket{b} =  \left| 
\begin{array} { @{\,}c@{\;}c@{\;}c@{\,}c@{\;}c@{\;}c@{\,}c@{\;}c@{\;}c@{\;}c@{\;}c@{\;}c@{}}
	\hat{I}_1 & \hat{I}_2 & {\bf S}_3^2 & \hat{I}_4 &  {\bf S}_5^2 & {\bf S}_6^2 &  \hat{I}_7 & \hat{\bf J}_{12}^2 & \hat{\bf J}_{34}^2 & \hat{\bf J}_{125}^2 &  \hat{\bf J}_{1256}^2 &  \hat{\bf J}_{{\rm tot}}  \\
	j_1 & j_2 & s_3 & j_4 & s_5 & s_6 & j_7 & j_{12} & j_{34} & j_{125} & j_{1256} & {\bf 0} 
\end{array}  \right>  \, .
\label{ch8: eq_15j_3_12_b_state}
\end{equation}
In the above notation, the large ket lists the operators on the top row, and the corresponding quantum numbers are listed on the bottom row. See Eq.\ (8) - (12) in \cite{yu2011a} for the definitions of some of the operators listed in Eq.\ (\ref{ch8: eq_15j_3_12_a_state}) and (\ref{ch8: eq_15j_3_12_b_state}). The ones that require new definitions are

\begin{eqnarray}
({\hat{J}}_{135}^2)_{\alpha \beta}  &=& [  J_{1}^2 + \hbar^2 s_3(s_3+1) + \hbar^2 s_5(s_5+1) ] \delta_{\alpha \beta} 
+ 2 {\bf \hat{J}}_{1} \cdot [ ({\bf S}_3)_{\alpha \beta} +  ({\bf S}_5)_{\alpha \beta} ] ,  \nonumber  \\  
 &&  \\
({\hat{J}}_{1356}^2)_{\alpha \beta}  &=& [  J_{1}^2 + \hbar^2 s_3(s_3+1) + \hbar^2 s_5(s_5+1) + \hbar^2 s_6(s_6+1) ] \delta_{\alpha \beta}   \nonumber  \\
&& \quad \quad \quad \quad \quad \quad  \quad \quad \quad \quad + 2 {\bf \hat{J}}_{1} \cdot [ ({\bf S}_3)_{\alpha \beta} +  ({\bf S}_5)_{\alpha \beta} +  ({\bf S}_6)_{\alpha \beta} ] ,    
  \\   \nonumber   \\
 ({\hat{J}}_{125}^2)_{\alpha \beta} &=& [  J_{12}^2+ \hbar^2 s_5(s_5+1) ] \delta_{\alpha \beta} 
 + 2 {\bf \hat{J}}_{12} \cdot ({\bf S}_5)_{\alpha \beta}  ,   \\  
({\hat{J}}_{1256}^2)_{\alpha \beta} &=& [  J_{12}^2+ \hbar^2 s_5(s_5+1) + \hbar^2 s_6(s_6+1) ] \delta_{\alpha \beta} 
 + 2 {\bf \hat{J}}_{12} \cdot [ ({\bf S}_5)_{\alpha \beta} + ({\bf S}_6)_{\alpha \beta} ] ,  \nonumber  \\  
 &&  \\
\label{ch8: eq_15j_3_12_Jtot_vector}
({\bf \hat{J}}_{{\rm tot}})_{\alpha \beta}  &=& ( {\bf \hat{J}}_1 + {\bf \hat{J}}_2 + {\bf \hat{J}}_4 + {\bf \hat{J}}_7 ) \delta_{\alpha \beta} + \hbar \, ({\bf S}_3)_{\alpha \beta} + \hbar \, ({\bf S}_5)_{\alpha \beta} + \hbar \, ({\bf S}_6)_{\alpha \beta}.  
\end{eqnarray}

By comparing Eq.\ (13) in \cite{yu2011a} with Eq.\ (\ref{ch8: eq_15j_3_12_Jtot_vector}) above, we find that the role of ${\bf \hat{J}}_5$ in Ref.\ \cite{yu2011b} is played by ${\bf \hat{J}}_7$ here. Otherwise, the operators associated with the large quantum numbers are exactly the same. The addition of the spinors ${\bf S}_5$ and ${\bf S}_6$ will introduce extra spinor products in the form of two additional $d$-matrices, as well as additional correction terms to the phase in the asymptotic formula. 

Following the derivations for the $9j$-symbol with one small quantum number in \cite{yu2011a}, we have derived  an asymptotic formula for the $15j$-symbol with three small quantum numbers. The result is

\begin{eqnarray}
\label{ch8: eq_main_formula_15j_3_12}
&&  \left\{
  \begin{array}{ccccc}
    j_1 & j_2 & j_{12} & j_{125} & j_{1256}\\ 
    s_3 & j_4 & j_{34} &  j_{135} & j_{1356} \\ 
    j_{13} & j_{24} & s_5 & s_6 & j_7 \\
  \end{array} 
  \right\}    \\   \nonumber
&=&  
(-1)^{ j_1 + j_2 + j_4 + j_7 + 2s_3 + \nu_3 + \mu_5 + \mu_6} \; \frac{ d^{s_3}_{\nu_3 \, \mu_3} (\theta_1) \; d^{s_5}_{\nu_5 \, \mu_5} (\theta_2) \; d^{s_6}_{\nu_6 \, \mu_6} (\theta_2)}{\sqrt{ [j_{34}][j_{13}][j_{135}] [j_{1356}][j_{125}][j_{1256}]  (12 \pi V) }} \,  \\   \nonumber
&&  \cos \left(  \sum_i  \, (j_i+\frac{1}{2}) \, \psi_i  + \frac{\pi}{4} - s_3  \pi  
 + \mu_3  \phi_4'  + \nu_3  \phi_1'  - (\mu_5 + \mu_6) \phi_{12} - (\nu_5 + \nu_6) \phi_1 \right)   \, .
\end{eqnarray}
The formula above is similar to Eq.\ (1) in \cite{yu2011a}, except for the two additional $d$-matrices associated with ${\bf S}_5$ and ${\bf S}_6$, and the extra phase corrections proportional to $\mu_5$, $\nu_5$, $\mu_6$ and $\nu_6$ in the argument of the cosine.

\begin{figure}[tbhp]
\begin{center}
\includegraphics[width=0.70\textwidth]{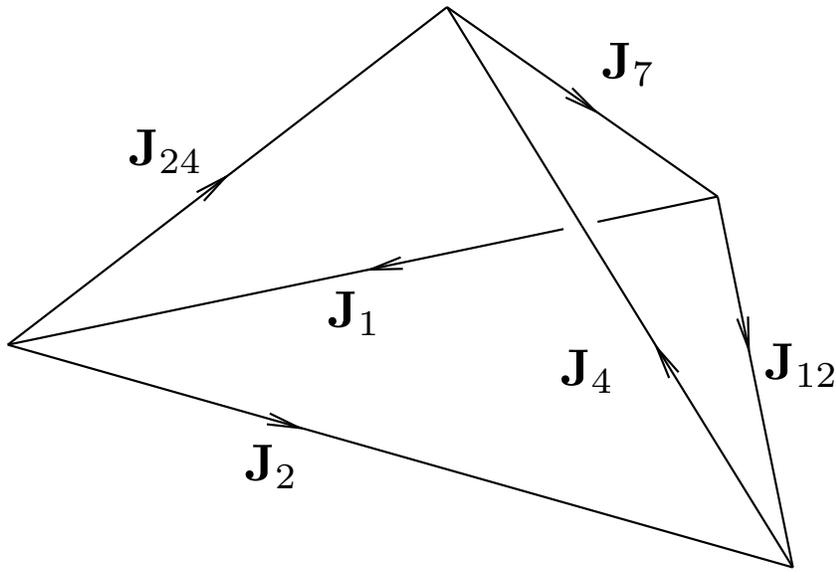}
\caption{The volume $V$ and the external dihedral angles $\psi_i$ are defined on the tetrahedron with the six edge lengths $J_1, J_2, J_4, J_7, J_{12}, J_{24}$. }
\label{ch8: fig_tetrahedra_15j_3_12}
\end{center}
\end{figure}

In Eq.\ (\ref{ch8: eq_main_formula_15j_3_12}), the indices for the $d$-matrices are $\mu_3 = j_{34}-j_4$, $\nu_3 = j_{13}-j_1$, $\mu_5 = j_{125}-j_{12}$, $\nu_5 = j_{135}-j_{13}$, $\mu_6 = j_{1256} - j_{125}$, and $\nu_6 = j_{1356} - j_{135}$. The first sum in the argument of the cosine is the Ponzano-Regge phase \cite{ponzano-regge-1968}, where $\psi_i$ are the six external dihedral angles in the tetrahedron with the six edge lengths $J_i$, $i = 1,2,4,7, 12, 24$. This tetrahedron is illustrated in Fig.\ \ref{ch8: fig_tetrahedra_15j_3_12}. See Ref.\ \cite{yu2011a} for more information on the construction of this tetrahedron.

The angles $\phi_1$ and $\phi_{12}$ are internal dihedral angles at the edge $J_1$ and $J_{12}$, respectively, in the tetrahedron in Fig.\ \ref{ch8: fig_tetrahedra_15j_3_12}. In other words, $\phi_1 = \pi - \psi_1$ and $\phi_{12} = \pi - \psi_{12}$.  The angle $\phi_1'$ is the angle between the $({\bf J}_1, {\bf J}_4)$ plane and the $({\bf J}_1, {\bf J}_{24})$ plane. The angle $\phi_4'$ is the angle between the $({\bf J}_1, {\bf J}_4)$ plane and the $({\bf J}_4, {\bf J}_{12})$ plane. Here we put primes on these angles to distinguish them from the internal dihedral angles $\phi_1$ and $\phi_4$. The angle $\theta_1$ is the angle between ${\bf J}_1$ and ${\bf J}_4$. The angle $\theta_2$ is the angle between ${\bf J}_1$ and ${\bf J}_{12}$. Explicitly, the angles $\phi_1', \phi_4'$, $\theta_1$, and $\theta_2$ are given by the following equations:

\begin{eqnarray}
\label{ch8: eq_phi_1p_def}
\phi_{1}' &=&  \pi - \cos^{-1} \left( \frac{ ({\bf J}_{1} \times {\bf J}_{4} ) \cdot ({\bf J}_{1} \times {\bf J}_{7} ) }{ | {\bf J}_{1} \times {\bf J}_{4} | \,  | {\bf J}_{1} \times {\bf J}_{7}  |} \right) \, ,    \\
\label{ch8: eq_phi_4p_def}
\phi_{4}' &=& \pi -  \cos^{-1} \left(  \frac{ ({\bf J}_{4} \times {\bf J}_{1} ) \cdot ({\bf J}_{4} \times {\bf J}_{7} ) }{ | {\bf J}_{4} \times {\bf J}_{1} | \,  | {\bf J}_{4} \times {\bf J}_{7}  |}  \right)  \, ,   \\
\label{ch8: eq_theta_15_3_12_def}
\theta_1 &=& \cos^{-1} \left( \frac{ {\bf J}_{1} \cdot {\bf J}_{4} }{J_{1} J_{4} }  \right)  \, ,  \\
\theta_2 &=& \cos^{-1} \left( \frac{ {\bf J}_{1} \cdot {\bf J}_{12} }{J_{1} J_{12} }  \right)  \, .
\end{eqnarray}

We plot the exact values of the $15j$-symbol against our approximation (\ref{ch8: eq_main_formula_15j_3_12}) for the following values of the $j$'s: 

\begin{equation}
\label{ch8: eq_15j_values_3_12}
 \left\{
  \begin{array}{ccccc}
    j_1 & j_2 & j_{12} & j_{125} & j_{1256} \\ 
    s_3 & j_4 & j_{34} &  j_{135}  & j_{1356} \\ 
    j_{13} & j_{24} & s_5 & s_6  & j_7  \\
  \end{array} 
  \right\}   
= 
	\left\{
  \begin{array}{rrrrr}
    203/2 & 207/2 & 96 & 97 & 98 \\ 
     3/2 & 199/2 & 100 & 100 & 101  \\ 
    101 & 108 & 1 & 1 & j_7 \\
  \end{array} 
  \right\}  \, . 
\end{equation}
Wee see that there are generally good agreements.

\begin{figure}[tbhp]
\begin{center}
\includegraphics[width=0.80\textwidth]{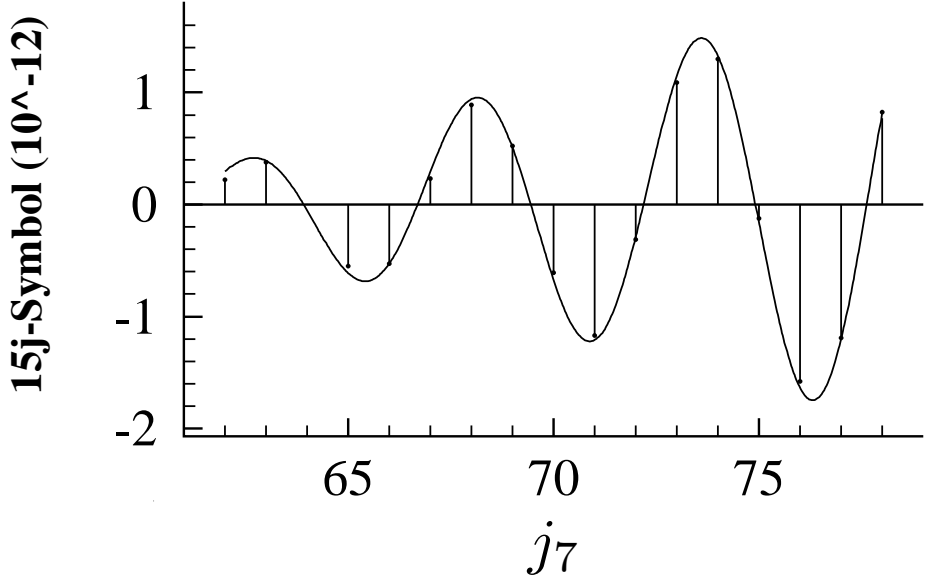}
\caption{Comparison of the exact $15j$-symbol (vertical sticks and dots) and the asymptotic formula (\ref{ch8: eq_main_formula_15j_3_12}), for the values of $j$'s shown in (\ref{ch8: eq_15j_values_3_12}). }
\label{ch8: fig_plot_15j_3_12}
\end{center}
\end{figure}

\newpage

\section{\label{ch8: sec_15j_4_11_case}Four Small Quantum Numbers}

We now take four angular momenta to be small.  We choose $j_1=s_1$, $j_4=s_4$, $j_5=s_5$, and $j_6=s_6$ to be small. Expressing the $15j$-symbol as an inner product of two multicomponent wave-functions, we have

\begin{eqnarray}
&& \left\{
   \begin{array}{ccccc}
    s_1 & j_2 & j_{12} & j_{125} & j_{1256} \\ 
    j_3 & s_4 & j_{34} &  j_{135}  & j_{1356} \\ 
    j_{13} & j_{24} & s_5 & s_6  & j_7  \\
  \end{array} 
  \right \}     \\  \nonumber
&=& \frac{\braket{ b  |  a } }{ \{ [j_{12}][j_{34}][j_{13}][j_{24}] [j_{125}] [j_{135}] [j_{1256}] [j_{1356}] \}^{\frac{1}{2}}}  \, , 
\end{eqnarray}
where

\begin{equation}
\ket{a} =  \left| 
\begin{array} { @{\,}c@{\;}c@{\;}c@{\,}c@{\;}c@{\;}c@{\,}c@{\;}c@{\;}c@{\;}c@{\;}c@{\;}c@{}}
	{\bf S}_1^2 & \hat{I}_2 & \hat{I}_3 & {\bf S}_4^2 &  {\bf S}_5^2 & {\bf S}_6^2 &  \hat{I}_7 & \hat{\bf J}_{13}^2 & \hat{\bf J}_{24}^2 & \hat{\bf J}_{135}^2 &  \hat{\bf J}_{1356}^2 &  \hat{\bf J}_{{\rm tot}}  \\
	s_1 & j_2 & j_3 & s_4 & s_5 & s_6 & j_7 & j_{13} & j_{24} & j_{135} & j_{1356} & {\bf 0} 
\end{array}  \right>  \, ,
\label{ch8: eq_15j_4_11_a_state}
\end{equation}

\begin{equation}
\ket{b} =  \left| 
\begin{array} { @{\,}c@{\;}c@{\;}c@{\,}c@{\;}c@{\;}c@{\,}c@{\;}c@{\;}c@{\;}c@{\;}c@{\;}c@{}}
	{\bf S}_1^2 & \hat{I}_2 & \hat{I}_3 & {\bf S}_4^2 &  {\bf S}_5^2 & {\bf S}_6^2 &  \hat{I}_7 & \hat{\bf J}_{12}^2 & \hat{\bf J}_{34}^2 & \hat{\bf J}_{125}^2 &  \hat{\bf J}_{1256}^2 &  \hat{\bf J}_{{\rm tot}}  \\
	s_1 & j_2 & j_3 & s_4 & s_5 & s_6 & j_7 & j_{12} & j_{34} & j_{125} & j_{1256} & {\bf 0} 
\end{array}  \right>  \, .
\label{ch8: eq_15j_4_11_b_state}
\end{equation}
The states in Eq.\ (\ref{ch8: eq_15j_4_11_a_state}) and (\ref{ch8: eq_15j_4_11_b_state}) are analogous to the states in Eq.\ (A2) and (A3) in Ref.\ \cite{yu2011a}, except that the role of ${\bf \hat{J}}_5$ in Ref.\ \cite{yu2011a} is played by ${\bf \hat{J}}_7$ here. There are also two additional spinors ${\bf S}_5$ and ${\bf S}_6$, which will introduce two extra spinor products in the form of two $d$-matrices in the final asymptotic formula. 

\begin{figure}[tbhp]
\begin{center}
\includegraphics[width=0.55\textwidth]{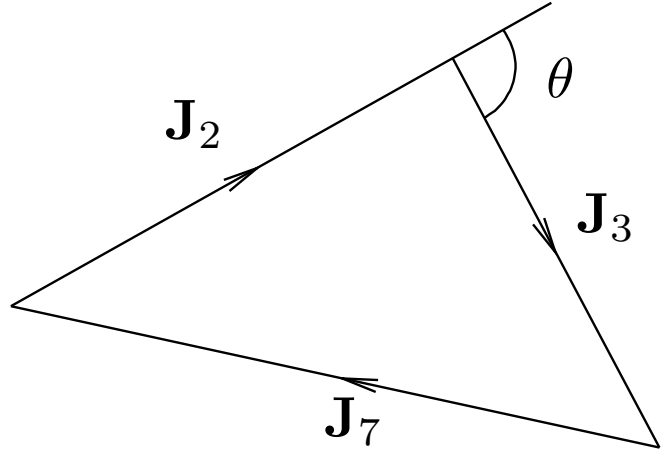}
\caption{The angle $\theta$ is the exterior angle between the edges $J_2$ and $J_3$ in a triangle having the three edge lengths $J_2, J_3, J_7$.}
\label{ch8: fig_15j_4_11_triangle}
\end{center}
\end{figure}

Following the derivations for the $9j$-symbol with two small quantum number in Appendix A in \cite{yu2011a}, we have derived  an asymptotic formula for the $15j$-symbol with two small quantum numbers. The result is

\begin{eqnarray}
\label{ch8: eq_formula_15j_4_11}
&&   \left\{
  \begin{array}{ccccc}
    s_1 & j_2 & j_{12} & j_{125} & j_{1256} \\ 
    j_3 & s_4 & j_{34} & j_{135} & j_{1356} \\ 
    j_{13} & j_{24} & s_5 & s_6  & j_{7} \\
  \end{array} 
  \right\} \\   \nonumber
	&=& \frac{(-1)^{j_2 + j_3 + j_7 + \mu_1 + \mu_4 + \mu_5 + \mu_6}}{\sqrt{[j_{12}][j_{34}][j_{13}][j_{24}][j_{125}][j_{135}][j_{1256}][j_{1356}]}}   
	 \, d^{(s_1)}_{\nu_1 \, \mu_1} (\theta) \, d^{(s_4)}_{\nu_4 \, \mu_4} (\theta)  \, d^{(s_5)}_{\nu_5 \, \mu_5} (\theta) \, d^{(s_6)}_{\nu_6 \, \mu_6} (\theta) \, .
\end{eqnarray}
The indices on the $d$-matrices are $\mu_1 = j_{12}-j_2$, $\nu_1 = j_{13}-j_3$,  $\mu_4 = j_{24}-j_2$, $\nu_4 = j_{34}-j_3$,  $\mu_5 = j_{125}-j_{12}$, $\nu_5 = j_{135}-j_{13}$,  $\mu_6 = j_{1256}-j_{125}$, $\nu_6 = j_{1356}-j_{135}$. The angle $\theta$ is the angle between the vectors $\vec{J}_2$ and $\vec{J}_3$ in the triangle with the three edge lengths $J_2$, $J_3$, and $J_7$. This triangle is illustrated in Fig.\ \ref{ch8: fig_15j_4_11_triangle}. Explicitly, The angle $\theta$ is given by
\begin{equation}
	\theta = \cos^{-1} \left( \frac{{\mathbf J}_2 \cdot {\mathbf J}_3}{J_2 \, J_3} \right) = \pi - \cos^{-1} \left( \frac{ J_2^2 + J_3^2 - J_7^2 }{2 \, J_2 \, J_3 } \right) \, . 
\end{equation}

\begin{figure}[tbhp]
\begin{center}
\includegraphics[width=0.85\textwidth]{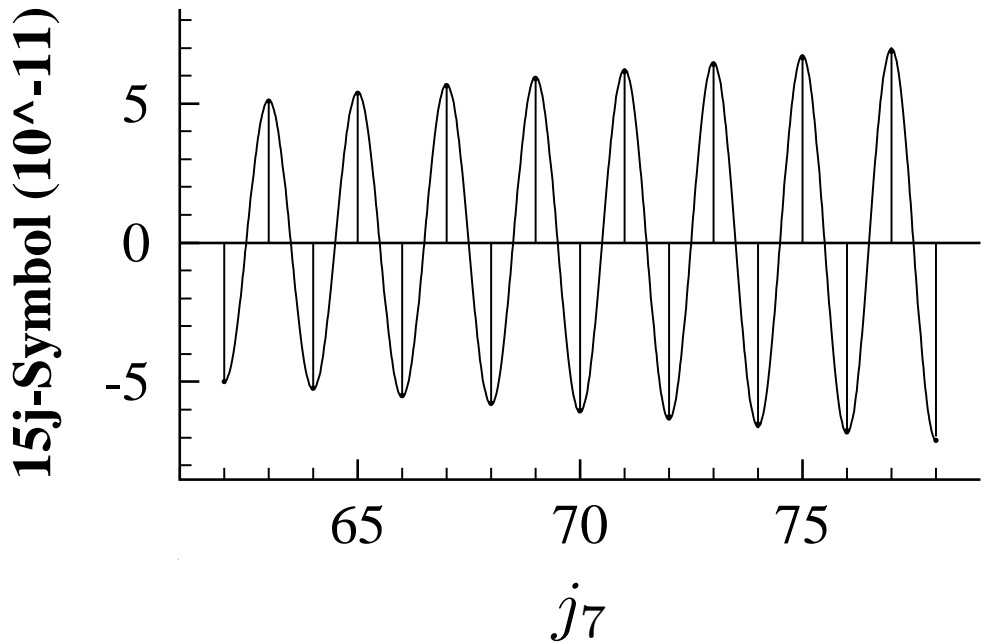}
\caption{Comparison of the exact $15j$-symbol (vertical sticks and dots) and the asymptotic formula (\ref{ch8: eq_formula_15j_4_11}), for the values of $j$'s shown in (\ref{ch8: eq_15j_values_4_11}). }
\label{ch8: fig_plot_15j_4_11}
\end{center}
\end{figure}

We illustrate the accuracy of the approximation (\ref{ch8: eq_formula_15j_4_11}) by plotting it against the exact $15j$-symbol in Fig.\ \ref{ch8: fig_plot_15j_4_11}  for the following values of the $j$'s:

\begin{equation}
\label{ch8: eq_15j_values_4_11}
 \left\{
  \begin{array}{ccccc}
    s_1 & j_2 & j_{12} & j_{125} & j_{1256} \\ 
    j_3 & s_4 & j_{34} &  j_{135}  & j_{1356} \\ 
    j_{13} & j_{24} & s_5 & s_6  & j_7  \\
  \end{array} 
  \right\}   
= 
	\left\{
  \begin{array}{rrrrr}
     1/2 & 237/2  & 118 & 119 & 118 \\ 
     189/2 & 3/2 & 94 & 94 & 95  \\ 
     95 & 117 & 1 & 1 & j_7 \\
  \end{array} 
  \right\}  \, . 
\end{equation}
There is generally good agreement.

\pagebreak

\section{Comments and Conclusions}

In this paper, we have presented three new asymptotic formulas for the Wigner $15j$-symbol, in the limit of a mixture of small and large quantum numbers. These results do not cover all the different asymptotic limits of the $15j$-symbol. More work in this area is needed. For example, base on the general method developed in \cite{yu2011a, yu2011b}, the asymptotic limit of the $15j$-symbol with one small angular momentum will be based on the yet unknown semiclassical analysis of the $12j$-symbol with all quantum numbers taken large. 

Since the $15j$-symbol is related to the $4$-simplex in four-dimensional quantum gravity, while the $6j$-symbol is related to the $3$-simplex in three-dimensional quantum gravity, the fact that the asymptotic formula for the $15j$-symbol with three small quantum numbers is expressed in terms of the tetrahedron of a $6j$-symbol is interesting. However, some of the extra angles, such as $\phi_1'$ and $\phi_4'$, that modify the Ponzano-Regge phase are not dihedral angles of quantized tetrahedra, so it is difficult to interpret the asymptotic formula in terms of the discrete Regge curvature. 

In this paper, we have inserted a particular set of $SU(2)$ intertwiners symmetrically into the pentagon diagram to obtain a spin network, which turns out to be a $15j$-symbol of the first kind. We can insert another set of $SU(2)$ intertwiners symmetrically into the pentagon diagram to obtain the $15j$-symbols of the fifth kind \cite{yutsis1962}. It will be interesting to apply the methods here to derive asymptotic formulas for this other kind of $15j$-symbol.

\section*{References}
\bibliographystyle{plain}
\bibliography{SLQN15J}

\end{document}